\documentclass{optica-article}

\journal{oe}

\articletype{Research Article}
\usepackage{makecell}
\usepackage{lineno}

\begin{document}

\title{Comparison of multi-mode Hong-Ou-Mandel interference and multi-slit interference}

\author{Yan Guo,\authormark{1,3} Zi-Xiang Yang,\authormark{1,3} Zi-Qi Zeng,\authormark{1} Chunling Ding,\authormark{1} Ryosuke Shimizu,\authormark{2}and Rui-Bo Jin\authormark{1,*}}

\address{\authormark{1}Hubei Key Laboratory of Optical Information and  Pattern Recognition, Wuhan Institute of Technology, Wuhan 430205, China\\
\authormark{2}The University of Electro-Communications, 1-5-1 Chofugaoka, Chofu, Tokyo, Japan\\
\authormark{3}These authors contributed equally to this work
}

\email{\authormark{*}Corresponding author: jin@wit.edu.cn} 
\homepage{http://www.qubob.com} 
\homepage{http://rs.pc.uec.ac.jp}


\begin{abstract}
Hong-Ou-Mandel (HOM) interference of multi-mode frequency entangled states plays a crucial role in  quantum metrology.
However, as the number of modes increases, the HOM interference pattern becomes increasingly complex, making it challenging to comprehend intuitively.
To overcome this problem, we present the theory and simulation of multi-mode-HOM interference (MM-HOMI)  and compare it to multi-slit interference (MSI).
We find that these two interferences have a strong mapping relationship and are determined by two factors:  the envelope factor and the details factor. The envelope factor is contributed by the single-mode HOM interference (single-slit diffraction)  for MM-HOMI (MSI).  The details factor is given by $\sin(Nx)/ \sin(x)$ ($[\sin(Nv)/\sin(v)]^2$) for MM-HOMI (MSI), where $N$ is the mode (slit) number and  $x (v)$ is the phase spacing of two adjacent spectral modes (slits).
As a potential application, we demonstrate that the square root of the maximal Fisher information in  MM-HOMI increases linearly with the number of modes,   indicating that MM-HOMI is a powerful tool for enhancing precision in time estimation.
We also discuss multi-mode Mach–Zehnder interference,   multi-mode NOON-state interference, and the extended Wiener-Khinchin theorem.
This work may provide an intuitive understanding of  MM-HOMI patterns and promote the application of MM-HOMI in quantum metrology. 
\end{abstract}

\section{Introduction}
Since its discovery in 1987, the  Hong-Ou-Mandel (HOM) interference using downconverted biphotons has shown a wide variety of applications in quantum optics  \cite{Hong1987, AgataM2017, Bouchard2020,Liu_2021,Yang_2022}. 
In traditional HOM interference, the biphotons are usually correlated in one discrete spectral mode. 
However, the biphotons involved can be correlated in multiple discrete spectral modes, and this two-body high-dimensional entangled state can be called entangled qudits \cite{Jin2016QST, Useche2021, Castro2022,Yang:23}.
Here, we define the HOM interference using frequency entangled qudits as the multi-mode HOM interference (MM-HOMI).
One important characteristic of  MM-HOMI is that its interference patterns are significantly narrower than those in single-mode HOM interference. 
Such narrow interference fringes provide more Fisher information in phase estimation \cite{Xiang2013, Jin2016SR, Lyons2018SA, Lingaraju2019, Chen2019njpQIursin,  Chen2021}. 
As a result,  MM-HOMI is very promising in quantum metrology.

Recently, many works have been devoted to the study of  HOM interference using biphotons in multi-frequency modes.
%
Lingaraju et al. investigated the effect of spectral phase coherence of multi-frequency modes in  HOM interference \cite{Lingaraju2019}.
%
Chen et al. utilized HOM interference as a tool to characterize up to six-mode frequency entangled qudits \cite{Chen2019njpQIursin, Chen2021}.
%
Morrison et al. prepared an eight-mode frequency entangled state in a customized poling crystal and tested its HOM interference patterns \cite{Morrison_2022}.
In addition, HOM interference using biphoton frequency combs, which have a large number of discrete frequency modes, has also been widely investigated \cite{Xue_2019, Fabre2020, Xie_2015,Chang_2021}.

However, as the mode number increases, the MM-HOMI pattern becomes more and more complicated, which makes it challenging to understand intuitively. 
To address this issue,  we first present the theory and simulation of MM-HOMI and then compare it with a well-known classical interference, the multi-slit interference (MSI)\cite{born1980, Young1807, Hariharan2003,Jewett2008, Young2012}.
We demonstrate that the MM-HOMI and the MSI exhibit a strong mapping relationship.

%

%
\graphicspath{ {images/} }
\begin{figure*}[htbp]
\centering
\includegraphics[width=0.85\textwidth]{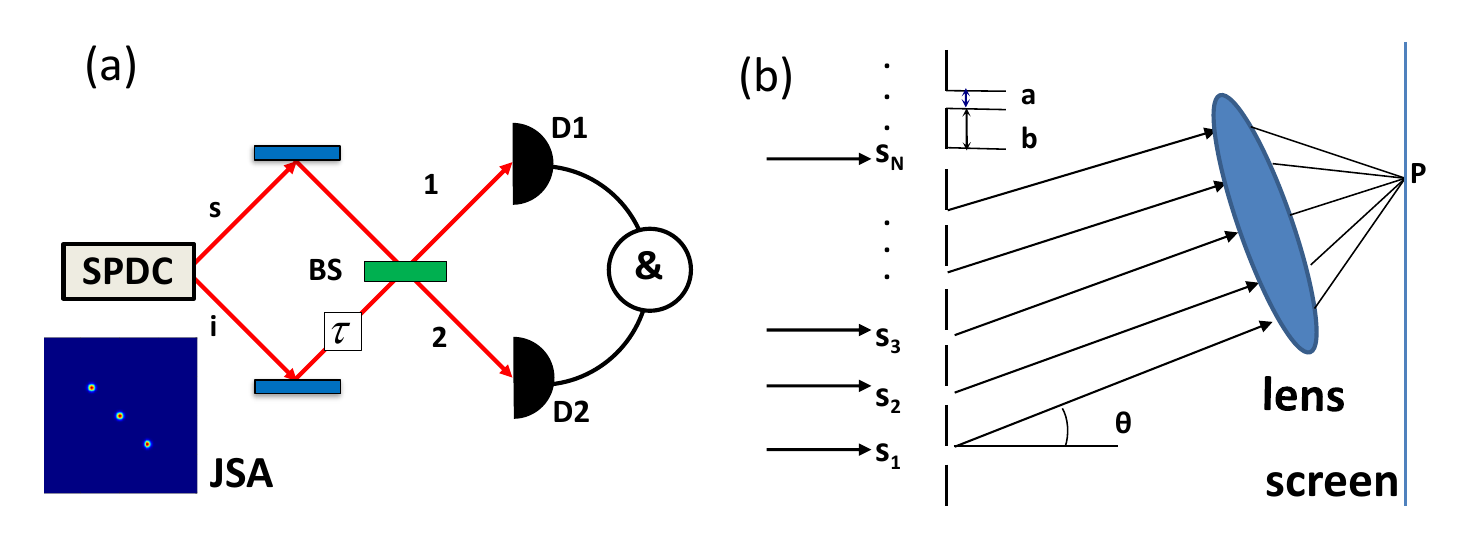}
\caption{ (a) The  typical setup  of multi-mode Hong-Ou-Mandel interference (MM-HOMI).
The signal (s) and idler (i) photons from an SPDC source impinge on a beam splitter (BS) before the signal is delayed by a time $\tau$. The output photons from the BS are detected by two single-photon detectors (D1 and D2), which are connected to a coincidence counter ($\&$). For an MM-HOMI, the biphotons have a multi-mode spectral distribution, as shown in the inset in the  bottom left corner. 
(b) The typical setup for a multi-slit interference. The  width of a single slit is $a$, and the width of a block is $b$; $a+b=d$. For point $p$, the tilt angle of the incident light is $\theta$.   The interference pattern is observed on the screen.
}
\label{fig:1}
\end{figure*}
%
%

\section{The theory and simulation of MM-HOMI}

The typical setup for HOM interference is shown in Fig.\,\ref{fig:1} (a).
The signal and idler photons  generated from a spontaneous parametric downconversion (SPDC) process can be expressed as \cite{URen2006, Mosley2008, LI2023OLT}
\begin{equation}\label{eq1}
\vert\psi\rangle= \iint \nolimits_{-\infty}^{+\infty}  d{\omega _s}d{\omega _i} f\left(\omega_s,\omega_i\right)\hat a_s^\dag\left(\omega_s\right)\hat a_i^\dag\left(\omega_i\right)\vert0\rangle\vert0\rangle,
\end{equation}
where $f\left(\omega_s,\omega_i\right) $ is the biphoton's joint spectral amplitude (JSA), $\omega$ is the angular frequency,  and ${\hat a^\dag }$ is the creation operator. The subscripts $s$ and $i$ represent the signal and idler photons, respectively.
In a HOM interference, the two-photon coincidence probability $P\left( \tau  \right)$ can be written as \cite{Grice1997,Jin2018Optica}
\begin{equation}\label{eq2}
\begin{array}{lll}
 P\left( \tau\right) = \frac{1}{2} - \frac{1}{2} \iint \nolimits_{-\infty}^{+\infty}  d{\omega _1}d{\omega _2}  {\left|f\left( {{\omega_1},{\omega _2}} \right)\right|^2}{\cos\left(\left( {{\omega _1} - {\omega _2}} \right)\tau\right) } ,  
\end{array}
 \end{equation}
 where $\omega _1$ and  $\omega _2$ are the frequencies detected by detectors D1 and D2 in Fig.\,\ref{fig:1} (a).
 For simplicity, in the above equation, we have assumed that $f\left(\omega_1,\omega_2\right)$ is  normalized and satisfies the exchanging symmetry of  $f\left(\omega_1,\omega_2\right)=f\left(\omega_2,\omega_1\right)$. See the Appendix for more details.

In the MM-HOMI,  the JSA can be written as: 
\begin{equation}\label{eq3}
\begin{array}{l}
f\left( {{\omega _1},{\omega _2}} \right) = \sum\limits_{k = 1}^N {{f_0}\left( {{\omega _1} - {\omega _0} - \left( {2k - N - 1} \right)\alpha ,{\omega _2} - {\omega _0} + \left( {2k - N - 1} \right)\alpha } \right)} ,\\ 
\end{array}
 \end{equation} 
where $f_0\left({\omega _1},{\omega _2}\right)$ is an arbitrary distribution function of the single spectral mode,  $N$ is the mode number,
$ \alpha $ represents the mode spacing, and $\omega _0$ is the mode's central frequency. 
As calculated in the Appendix,  $P\left( \tau  \right)$ can be simplified as
\begin{equation}\label{eq4}
\begin{array}{lll}
 P\left( \tau  \right) &= \frac{1}{2}\left[ 1-\frac{1}{{N}} \frac{{\sin (2N\alpha \tau )}}{{\sin (2\alpha \tau )}}\iint\nolimits_{ - \infty }^\infty  {d{\omega _1}}  d{\omega _2}{f_0}^2\left( {{\omega _1},{\omega _2}} \right)\cos \left( {\left( {{\omega _1} - {\omega _2}} \right)\tau } \right)\right]\\ 
  &= \frac{1}{2}\left[1 - P_0\frac{{{\rm{\sin}}\left( { N x } \right)}}{{{\rm{\sin}}\left( {x } \right)}} \right], \\
\end{array}
 \end{equation}
where $x=2 \alpha\tau$, which corresponds to the phase spacing caused by two adjacent spectral modes.  $P_0=\frac{1}{{N}}\iint\nolimits_{ - \infty }^\infty  {d{\omega _1}} d{\omega _2}{f_0}^2\left( {{\omega _1},{\omega _2}} \right)\cos \left( {\left( {{\omega _1} - {\omega _2}} \right)\tau } \right)$ 
corresponds to the envelope of the interference patterns.
We can observe in Eq.\,(\ref{eq4})  that the MM-HOMI is determined by two factors: the envelope factor $P_0$ and the details factor $\frac{{{\rm{\sin}}\left( { N x } \right)}}{{{\rm{\sin}}\left( {x } \right)}}$. $P_0$ is contributed by the single-mode HOM interference (for N=1). 
$P_0$ can also be expressed in the form of a Fourier transformation by projecting   ${f_0}^2\left( {{\omega _1},{\omega _2}} \right)$ on the axis of $\omega _1- \omega _2$,  as shown in Eq.\,(\ref{eq23}) in the Appendix.

For simplicity, we can set   $\;f\left( {{\omega _1},{\omega _2}} \right)$ to a multi-mode Gaussian distribution\cite{Morrison_2022, Zhu2023}: 
\begin{equation}\label{eq5}
\begin{array}{l}
f\left( {{\omega _1},{\omega _2}} \right) = \mathop \sum \limits_{k = 1}^N \exp \left[ - \frac{{{{({\omega _1} -{\omega _0}- \left( {2k - N - 1} \right)\alpha )}^2}}}{{{\gamma ^2}}} - \frac{{{{({\omega _2}-{\omega _0} + \left( {2k - N - 1} \right)\alpha )}^2}}}{{{\gamma ^2}}}\right],\\ 
\end{array}
 \end{equation} 
where $N$ is the mode number,  $ \gamma $ represents the mode width.
In a real experiment, $ \gamma $ and $\alpha$ are determined by the width of the pump and the phase-matching function of the crystal \cite{Morrison_2022}.  
As calculated in the Appendix,  $P_0\left( \tau  \right)$ can be simplified as
\begin{equation}\label{eq6}
\begin{array}{lll}
 P_0=\frac{1}{{N}}\exp \left( - \frac{{{\gamma ^2}{\tau ^2}}}{4}\right).
\end{array}
 \end{equation}

\graphicspath{ {images/} }
\begin{figure*}[htbp]
\centering
\includegraphics[width= 0.98\textwidth]{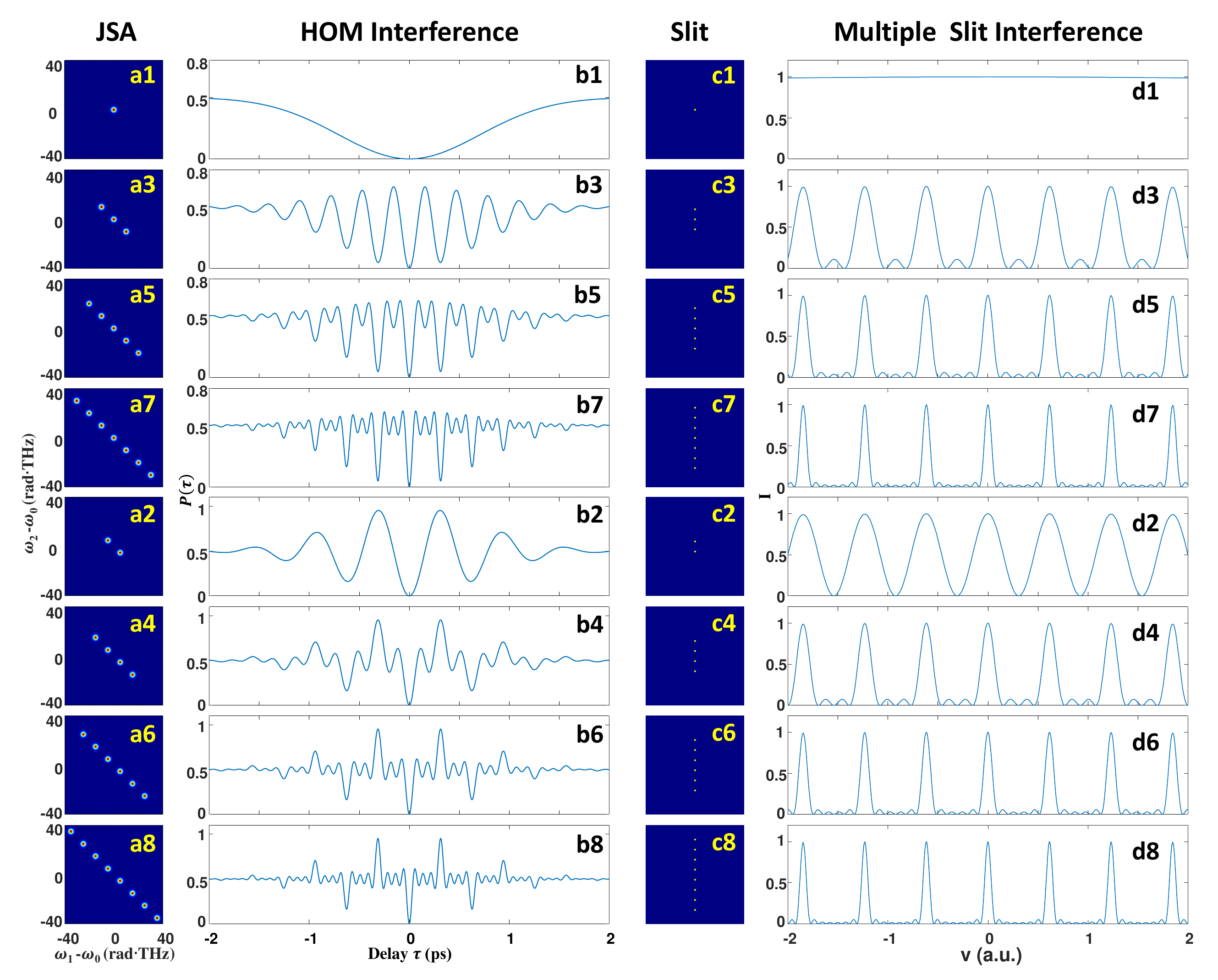}
\caption {Theoretical simulations of MM-HOMI and MSI. (a1-a8): The JSA of the biphotons with mode numbers ranging from 1 to 8,  with$\;\alpha $=5 rad$\cdot$THz and $\gamma = 2$ rad$\cdot$THz.
(b1-b8) : The MM-HOMI patterns based on the JSA shown in (a1-a8). 
(c1-c8): The slit distribution for $N=1-8$, with $a=1\times 10^{-5}$m and $d=5\times 10^{-4}$m.
(d1-d8): The MSI patterns based on the slit distribution in (c1-c8). 
}

\label{fig:2}
\end{figure*}
%

According to Eq.\,(\ref{eq4}), Eq.\,(\ref{eq5}) and Eq.\,(\ref{eq6}), we can plot the interference patterns of MM-HOMI.
As shown in Fig.\,\ref{fig:2}, the first column is the JSA of the biphotons, and the mode number is 1, 3, 5, 7, 2, 4, 6,  and 8.
Here, we choose the unit ellipticity for each mode, since this is the simplest case and has been realized in experiment \cite{Morrison_2022}.
The second column is the corresponding MM-HOMI pattern.

Firstly, we compare the envelopes of the interference patterns.
The odd-number MM-HOMI patterns have an asymmetric envelope, whereas the even-number MM-HOMI patterns are symmetric to the line of $P\left(\tau\right)=0.5$.
This is due to the characteristics of the function 
$\frac{\sin\left(N x\right)}{\sin\left(x\right)}$, which is asymmetric (symmetric) when N is an odd (even) number.

Secondly, we compare the primary and secondary valley (peak) numbers.
For the odd-number MM-HOMI patterns,  there are (N-3)/2  secondary valleys (for N$>$3) between two primary valleys, while for the even-number MM-HOMI patterns,  there are N-2  secondary valleys (for N$>$2) between two primary valleys.

\section{Comparison of  MM-HOMI and MSI}

So far, we have analyzed the properties of the MM-HOMI. 
However, we can notice  that the interference  patterns   in Fig.\,\ref{fig:2} are very complicated. 
One might gain some insight by comparing MM-HOMI with a classical well-known multi-slit interference (MSI).
It may be intuitive to understand that the multiple spectra function in a manner similar to the multiple slits.

Next, we deduce the mathematical form of MSI and compare it with  MM-HOMI.

The typical setup for MSI is shown in Fig.\,\ref{fig:1} (b). Here, the number of slits is $N$, the slit width is $a$,  the interval of the slits is $ b $, and $a+b=d$. 
As calculated in the Appendix, the amplitude  of the diffraction pattern at point $ p$ is
\begin{equation}\label{eq7}
\begin{array}{lll}
  A_p
  ={A_0}\frac{{\sin\left(u\right)}}{u}\frac{{\sin \left(Nv\right)}}{{\sin \left(v\right) }},
\end{array}
 \end{equation}
where 
$A_0$ is a constant  determined by the power of the light source, the distance between the slit and the screen, and the size of the slit \cite{born1980}.
$u = \frac{{\pi a\sin\theta }}{\lambda }$  represents the  phase difference in one slit.
$v = \frac{{\pi d\sin\theta }}{\lambda }  $ represents  the  phase difference between two adjacent slits.
$\lambda$ is the wavelength of the input light and $\theta$ is the tilt angle of the light, as shown in Fig.\,\ref{fig:1} (b).
For simplicity, we can set $A_0=1$.
The intensity of the diffraction pattern is:
\begin{equation}\label{eq8}
\begin{array}{lll}
I={\mid A_p\mid}^2=\left(\frac{\sin \left(u\right)}{u}\right)^2\left(\frac{\sin\left(Nv\right)}{\sin\left(v\right)}\right)^2
= I_0 \left(\frac{\sin\left(Nv\right)}{\sin\left(v\right)}\right)^2, \\
\end{array}
 \end{equation}
where ${I_0} =\left(\frac{{\sin}{\left(u\right)}}{u}\right)^2$ is the intensity due to one-slit diffraction, which is also the interference pattern's envelope.

According to Eq.\,(\ref{eq8}), we can plot the interference patterns of MSI, as shown in the third  and fourth columns in Fig.\,\ref{fig:2}.  The parameters are listed in detail in the caption of Fig.\,\ref{fig:2}.
There are N-2 secondary peaks (for N$>$2) between two primary peaks in the fourth column of Fig.\,\ref{fig:2}. This is comparable to the even-number MM-HOMI patterns,  which also have N-2  secondary valleys (for N$>$2) between two primary valleys.

\graphicspath{ {images/} }
\begin{figure*}[htbp]
\centering
\includegraphics[width= 0.98\textwidth]{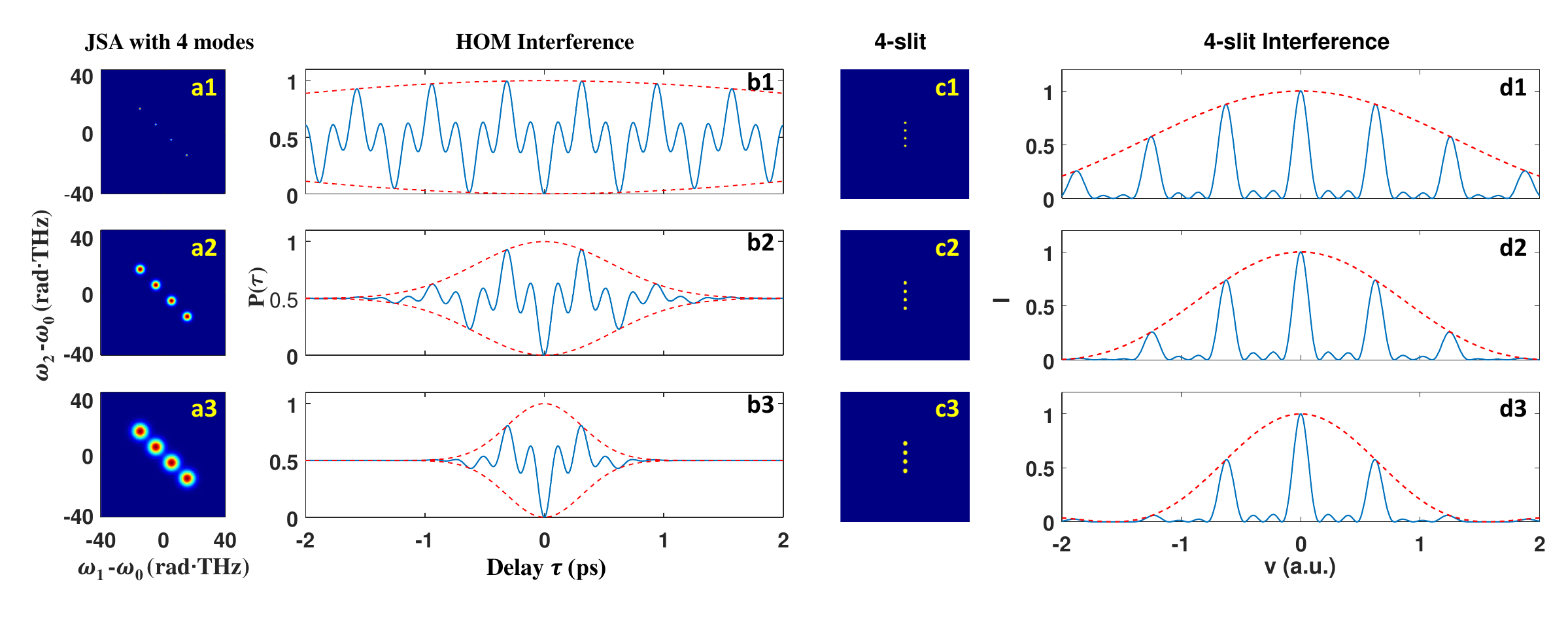}
\caption{(a1-a3): The JSA of the biphotons with mode spacing of  $\alpha $=5 rad$\cdot$THz and  mode sizes of $\gamma $= 0.5 rad$\cdot$THz, 2.5 rad$\cdot$THz, and 4.5 rad$\cdot$THz, respectively. 
(b1-b3): The calculated HOMI patterns using the JSA in (a1-a3). 
(c1-c3): The 4-slit distributions with  $d=5\times 10^{-4}$m and  $a=1\times 10^{-4}$m, $1.5\times 10^{-4}$m, and $2\times 10^{-4}$m, respectively.
(d1-d3): The calculated MSI patterns using the slits in (c1-c3). }
\label{fig:3}
\end{figure*}

Next, we investigate the influence of the mode size in MM-HOMI and MSI.
Figure\,\ref{fig:3} (a1-a3, b1-b3) displays the JSA of the biphotons and the corresponding MM-HOMI. Here, we set $\alpha $ to be fixed at 5 rad$\cdot$THz, and $\gamma$ to be 0.5 rad$\cdot$THz, 2.5 rad$\cdot$THz,  and 4.5 rad$\cdot$THz, respectively.
It can be observed that with the increase of $\gamma$, the envelope becomes narrower, but the spacing of adjacent peaks (valleys)  does not change.
This phenomenon can be well explained by Eq.\,(\ref{eq4}).
Figure\,\ref{fig:3} (c1-c3, d1-d3) shows the 4-slit distributions and the corresponding MSI. It can be noticed that the envelope also becomes narrower with the increase of $a$, which is similar to  the phenomenon of MM-HOMI.

\graphicspath{ {images/} }
\begin{figure*}[htbp]
\centering
\includegraphics[width= 0.98\textwidth]{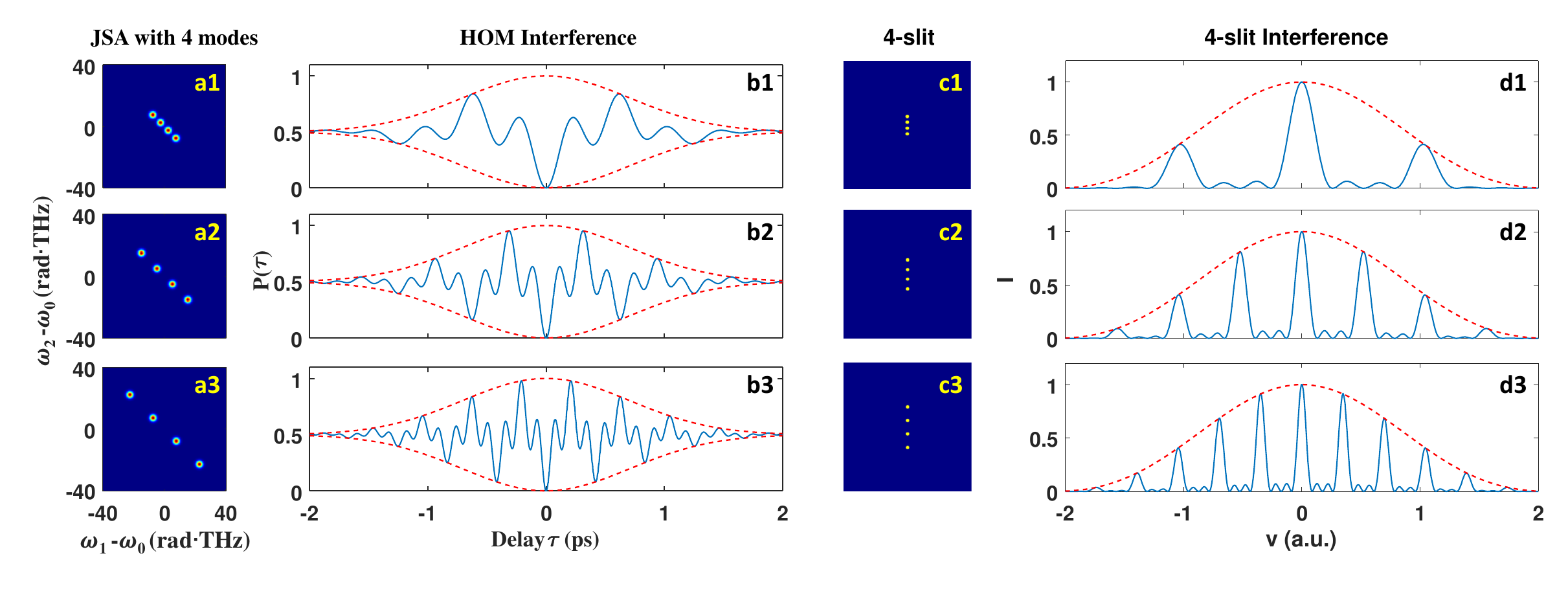}
\caption{
(a1-a3): The JSA of biphotons with mode size of $\gamma=2$ rad$\cdot$THz and mode spacing of $\alpha $=2.5 rad$\cdot$THz, 5 rad$\cdot$THz and 7.5 rad$\cdot$THz, respectively. 
(b1-b3): The calculated MM-HOMI patterns using the JSA in (a1-a3). 
(c1-c3): The 4-slit distributions with $a=1.5\times 10^{-4}$m, and $d=3\times 10^{-4}$m, $6\times 10^{-4}$m, $9\times 10^{-4}$m, respectively. (d1-d3): The calculated MSI patterns using the slits in (c1-c3). 
}
\label{fig:4}
\end{figure*}

Then, let us examine the impact of mode spacing on  MM-HOMI and MSI.
Figure\,\ref{fig:4} (a1-a3, b1-b3) depicts the JSA and MM-HOMI of biphotons with  $\gamma$  fixed at   2 rad$\cdot$THz and $\alpha$ increasing from 2.5 rad$\cdot$THz to 5 rad$\cdot$THz and 7.5 rad$\cdot$THz.
Figure\,\ref{fig:4} (c1-c3, d1-d3) shows the slit distributions and the corresponding MSI.
By comparing (a1-b3) with (c1-d3), we can observe that the envelope remains constant, while the number of peaks and valleys  increases as mode spacing increases.
These phenomena can also be well explained by Eq.\,(\ref{eq4}) and  Eq.\,(\ref{eq8}).

After analyzing Figs.\,(\ref{fig:4}, \ref{fig:5}, \ref{fig:6})  and  Eqs.\,(\ref{eq4}, \ref{eq8}), we can confirm that the MM-HOMI and MSI have a  strong   mapping relationship, as summarized  in Tab.\,\ref{tab:1}.
The phase  variable $x$, which accumulates in the time domain, represents the phase spacing between two spectra modes in the MM-HOMI;
while the phase variable $v$, which  accumulates in the space domain, represents the phase spacing between two slits in the MSI.  
Both MM-HOMI and MSI are determined by two factors: the envelope factor and the details factor.
The details factor includes  a common term of $\frac{{{\rm{\sin}}\left( { N x } \right)}}{{{\rm{\sin}}\left( {x } \right)}}$ with a mode number of $N$.
In MM-HOMI, the envelope factor $P_0$ corresponds to the HOM interference of a single spectral mode, and $P_0$ is also related to the Fourier transform of a single spectral mode, as shown in Eq.\,(\ref{eq23}). 
Similarly, the details factor $I_0$ corresponds to the single-slit diffraction, and it is also contributed by the Fourier transform of a single slit, as explained in Eq.\,(\ref{eq28}) in the Appendix. 
Therefore, we can conclude that the multiple spectra indeed function similarly to the multiple slits, as we expected at the beginning of this section. Consequently, the mapping relationship really can help on the intuitive  understanding of the MM-HOMI.

\begin{table}[htbp]
\caption{The mapping relationship between MM-HOMI and MSI.} 
\centering
\begin{tabular}{ccc} 
\hline
  &\makecell{MM-HOMI\\$P= \frac{1}{2}\left[1 - P_0\frac{{{\rm{\sin}}\left( { N x } \right)}}{{{\rm{\sin}}\left( {x } \right)}}\right]$}& \makecell{MSI\\$I= I_0 \left(\frac{{\sin}\left(Nv\right)}{{\sin}{\left(v\right)}}\right)^2$}\\
\hline

variable &$x=2\alpha\tau$& $v = \frac{{\pi d\sin\theta }}{\lambda} $\\

domain &frequency ($\alpha$) $\leftrightarrow$ time ($\tau$) &
spatial frequency ($\frac{{ \sin\theta }}{\lambda }$)$\leftrightarrow$space ($d$) \\

phase & $x$: phase spacing of two  spectra & $v$: phase spacing of two slits\\

mode number& $N$ & $N$ \\

details factor& $\frac{\sin \left(Nx\right)}{\sin\left(x \right)}$ & $\left(\frac{\sin\left(Nv\right)}{\sin\left(v\right)}\right)^2$\\

envelope factor&  $P_0$: single-mode HOMI  & $I_0$: single-slit diffraction\\

Fourier transform &  $P_0$: FT of single spectral mode&  $I_0$: FT of single slit \\

\hline
\end{tabular}
\label{tab:1}
\end{table}

\section{Application of MM-HOMI in quantum meteorology}

MSI has numerous applications in optical measurement, with one typical example being the diffraction-grating-based spectrometer.
The resolving power of a diffraction grating is proportional to the total number of slits (or grooves)  on the grating \cite{born1980, Young1807, Hariharan2003,Jewett2008, Young2012}.
%
Inspired by this feature, here we consider the resolving power of a MM-HOMI in quantum metrology by increasing the  total number of  the spectral modes.

\begin{figure*}[htbp]
\centering
\includegraphics[width= 0.85\textwidth]{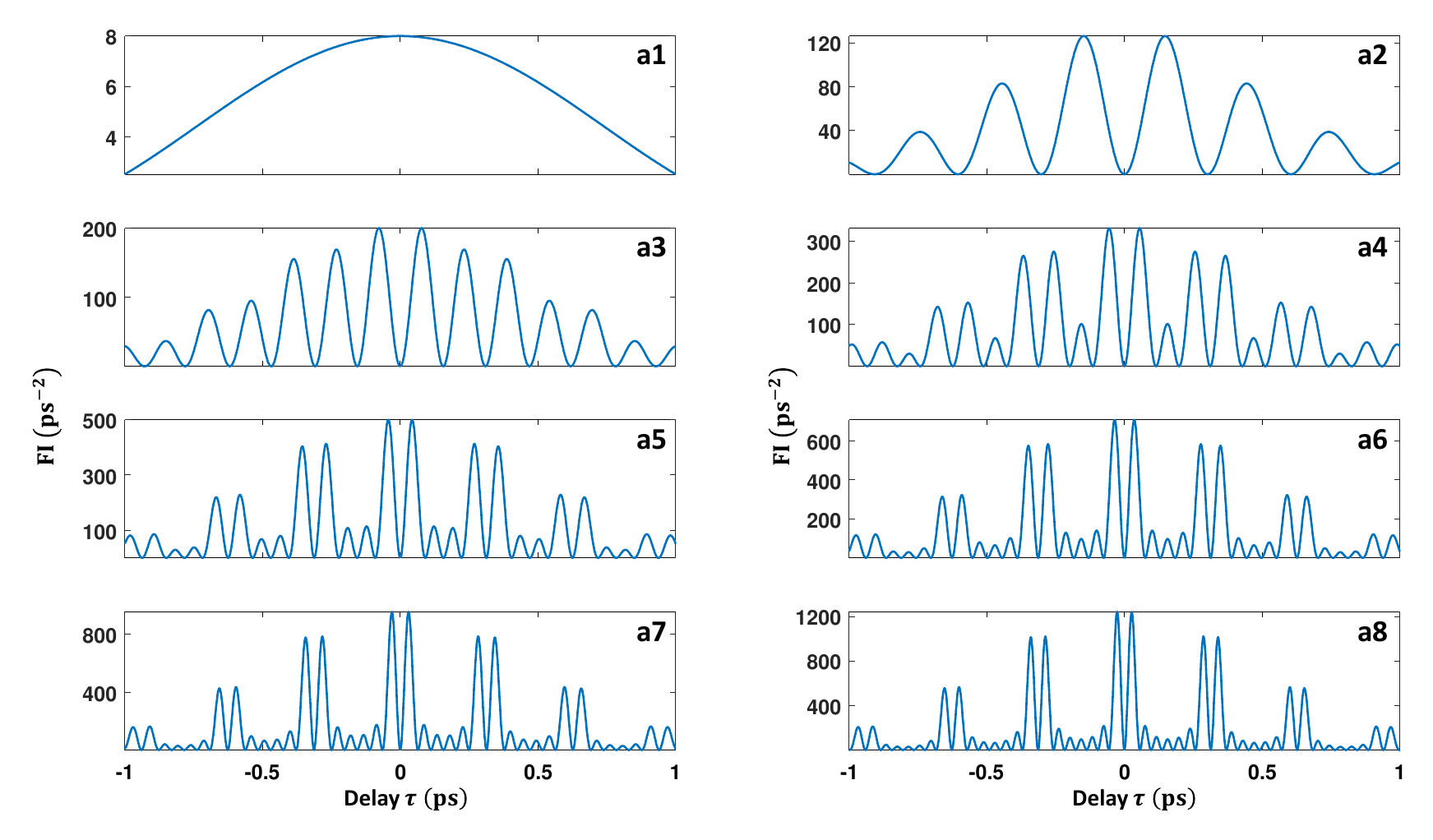}
\caption{ The Fisher information of MM-HOMI with a mode number ranging from 1 to 8 in (a1-a8),  with$\;\alpha $=5 rad$\cdot$THz and $\gamma = 2$ rad$\cdot$THz.  
 }
\label{fig:5}
\end{figure*}

\begin{figure*}[htbp]
\centering
\includegraphics[width= 0.75\textwidth]{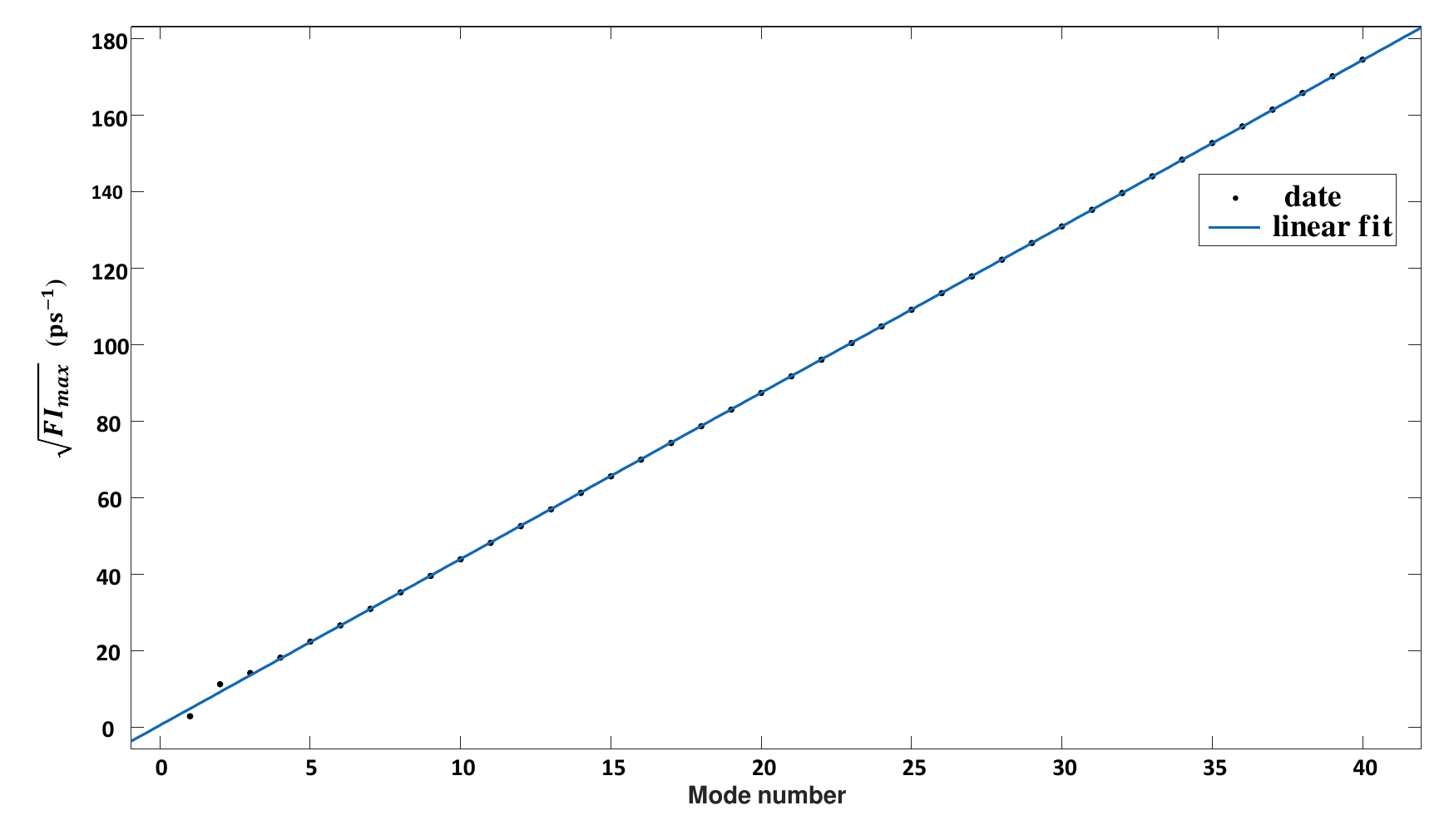}
\caption{ The square root of maximal Fisher information  with mode numbers ranging from 1 to 40, with$\;\alpha $=5 rad$\cdot$THz and $\gamma = 2$ rad$\cdot$THz. The blue line is a linear fit. }
\label{fig:6}
\end{figure*}

The ultimate limit on the precision of time estimation is the
Cram\'er-Rao bound \cite{Lyons2018SA,Cramer1999} , which states that the variance of any unbiased estimator is bounded by
\begin{equation}\label{eq9}
Var(\tilde t ) \ge \frac{1}{{Num \times FI}},
\end{equation}
where $\tilde t $ is the estimator of time $t$, and $Num$ is the number of measurement times and $FI$ is the Fisher information (FI).
For a single measurement, $Num$=1. So, the standard deviation (SD) is bound by 
\begin{equation}\label{eq10}
SD(\tilde t ) \ge \frac{1}{{\sqrt {FI}}}.
\end{equation}
FI of a single interference fringe can be calculated as  \cite{Xiang2013,Jin2021arXiv}:
\begin{equation}\label{eq11}
FI\left(\tau\right) = P\left(\tau \right)\left( {\frac{{\partial \left[\ln P\left(\tau \right)\right]}}{{\partial \tau }}} \right)^2  + \left[1 - P\left(\tau \right)\right]\left( {\frac{{\partial \left[\ln \left(1 - P\left(\tau \right)\right)\right]}}{{\partial \tau }}} \right)^2 = \frac{\left[ {P'\left(\tau \right)} \right]^2 }{{P\left(\tau \right)\left[1 - P\left(\tau \right)\right]}}.
\end{equation}
By using Eq.(\ref{eq4}) and Eq.(\ref{eq11}), we can obtain the Fisher information of MM-HOMI as 
\begin{equation}\label{eq12}
\begin{array}{lll}
FI\left( \tau  \right) =\frac{{{{\left( {\frac{{{\gamma ^2}\tau }}{2}\sin \left( {2\alpha N\tau } \right) - 2\alpha N\cos \left( {2\alpha N\tau } \right) + 2\alpha \sin \left( {2\alpha N\tau } \right)\cot \left( {2\alpha \tau } \right)} \right)}^2}}}{{{N^2}\exp [\frac{{{\gamma ^2}{\tau ^2}}}{2}]{{\sin }^2}\left( {2\alpha \tau } \right) - {{\sin }^2}\left( {2\alpha N\tau } \right)}}.
\end{array}
\end{equation}
Figure\,\ref{fig:5}(a1-a8) displays the simulated FI of the $P\left(\tau\right)$ in Fig.\,\ref{fig:2}(b1-b8), with the mode number N increasing from 1 to 8. 
The single valley in Fig.\,\ref{fig:2}(b8) is transferred to a double peak in Fig.\,\ref{fig:5}(a8).
Figure\,\ref{fig:6} summarizes the square root of maximal FI as a function of the  mode number N ranging from 1 to 40.
It is evident that the square root of maximal FI increases linearly with the increase in mode number.
This suggests that increasing the mode number  is a powerful method to improve precision in  time or phase estimation.

 \graphicspath{ {images/} }

\graphicspath{ {images/} }

\section{Discussion}
From the viewpoint of the extended Wiener-Khinchin theorem (e-WKT) \cite{Jin2018Optica}, the Fourier transform of the HOM interference  pattern is determined by the difference–frequency distribution of the JSI, i.e., the projection of  $f\left( {{\omega _1},{\omega _2}} \right)$ onto the $\omega _1 -\omega _2$ axis.
The e-WKT is not only  applicable to the single-mode case but also explains the multi-mode case in this study. 
Specifically, the interference patterns of MM-HOMI are determined by the Fourier transform of the difference–frequency distribution of the multi-mode JSI. 

\begin{figure}[!tb]
\centering
\includegraphics[width= 0.98\textwidth]{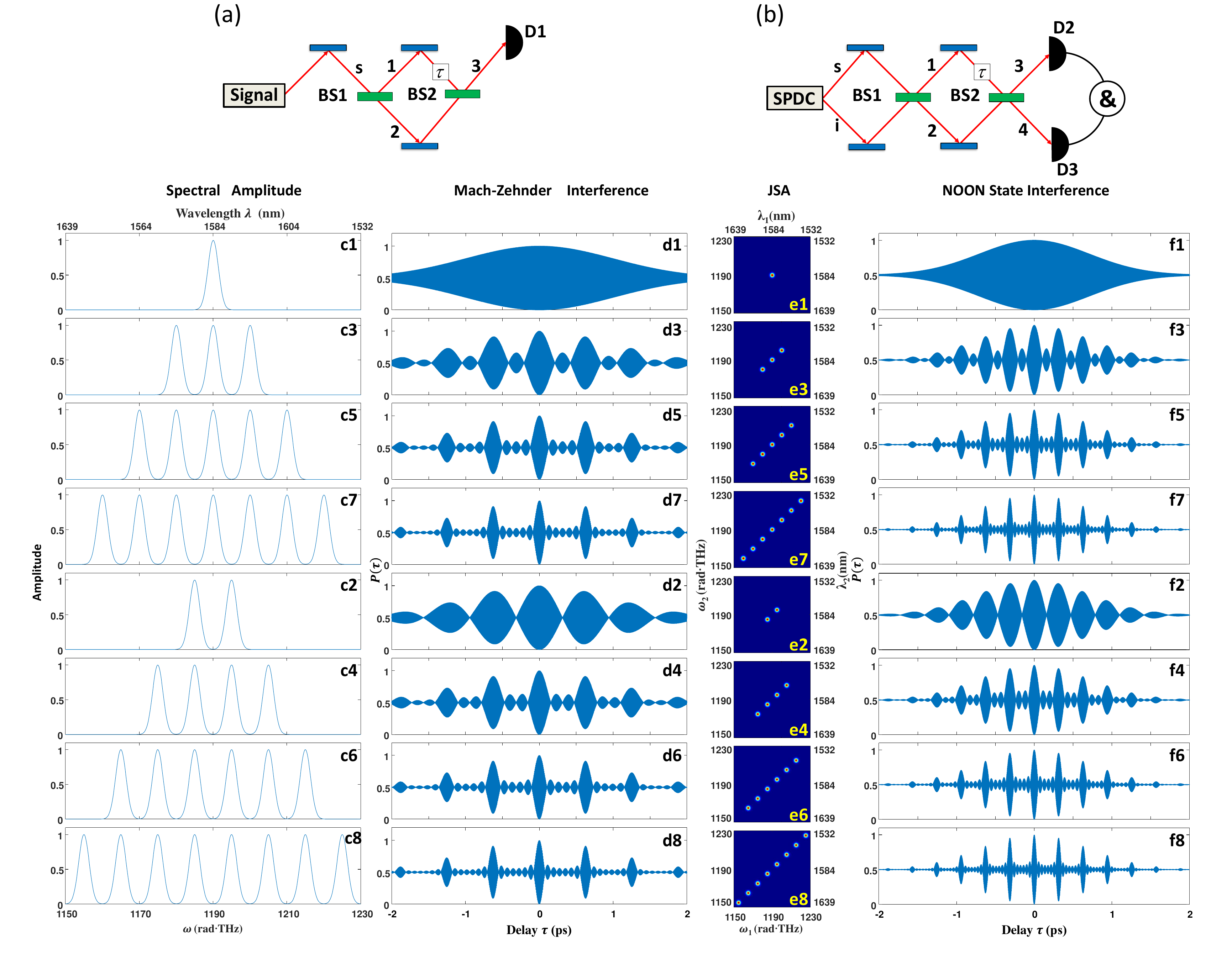}
\caption{
(a): The typical setup of Mach-Zehnder interference. 
(b): The typical setup of NOON-state interference.  (c1-c8): The spectral amplitude of the light source for multi-mode Mach–Zehnder interference.
(d1-d8): The multi-mode Mach–Zehnder interference patterns for mode numbers $N=1\sim8$  with$\;\alpha $=5 rad$\cdot$THz and $\gamma = 2$ rad$\cdot$THz.
(e1-e8): The joint spectral amplitude of the biphotons for the multi-mode NOON state interference.
(f1-f8): The multi-mode NOON state interference patterns for mode numbers $N=1-8$  with$\;\alpha $=5 rad$\cdot$THz and $\gamma = 2$ rad$\cdot$THz.
}
\label{fig:7}
\end{figure}

To gain a deeper understanding of the multi-mode effect, 
we also compared the MM-HOMI and MSI with  two other important interferences in quantum optics: the multi-mode Mach–Zehnder interference (MM-MZI) and the multi-mode NOON state interference (MM-NOONSI), using the setups shown in Fig.\,\ref{fig:7} (a, b).
 Here, the NOON-state is a 
$(\left| {20} \right\rangle  + \left| {02} \right\rangle )/\sqrt 2$ 
state, which has a photon number of 2, but with a spectral-mode number of N, as shown in  Fig.\,\ref{fig:7}(e1-e8).
As calculated in the Appendix, the single count in a MM-MZI can be expressed as 
\begin{equation}\label{eq13}
\begin{array}{l}
 P_{MZI}\left( \tau  \right)=
  \frac{1}{2} + \frac{1}{{2N}}\exp \left[ { - \frac{{{\gamma ^2}{\tau ^2}}}{8}} \right]\frac{{\sin \left( {N\alpha \tau } \right)}}{{\sin \left( {\alpha \tau } \right)}}\cos \left( {{\omega _0}\tau } \right),\\ 
\end{array}
 \end{equation}
 and the coincidence counts in a MM-NOONSI can be expressed as 
\begin{equation}\label{eq14}
\begin{array}{l}
 P_{NOON}\left( \tau  \right)=
  \frac{1}{2} + \frac{1}{{2N}}\exp \left[ - \frac{{{\gamma ^2}{\tau ^2}}}{4}\right]\frac{{\sin \left( {2N\alpha \tau } \right)}}{{\sin \left( {2\alpha \tau } \right)}}\cos \left( {2{\omega _0}\tau } \right).\\
\end{array}
 \end{equation}
In the above two models, Gaussian envelopes were chosen for simplicity. Refer to the Appendix for deductions using arbitrary envelopes.
The parameters are listed in detail in the caption of Fig.\,\ref{fig:7}.
By comparing the theoretical simulations in Fig.\,\ref{fig:7} (d1-d8) and (f1-f8), we observe that there are N-2 secondary peaks  (for N $>$ 3) between the two main peaks in N-mode  Mach–Zehnder interference  and N-mode  NOON-state interference.

By comparing Eq.\,(\ref{eq13}), Eq.\,(\ref{eq14}), Eq.\,(\ref{eq4}), and  Eq.\,(\ref{eq8}), we observe that the MM-MZI and MM-NOONSI are also contributed by the details factor of $\sin\left(Nx\right)/\sin\left(x\right)$, but multiplied by a factor of $\cos\left(\omega_0 \tau\right)$ or $\cos\left(2\omega_0 \tau\right)$.
In general, the connection between the MM-HOMI, the MSI, the MM-MZI, and the MM-NOONSI is that all these interferences are contributed by N input modes in physics and determined by the factor of  $\sin\left(Nx\right)/\sin\left(x\right)$ in mathematics.
%
\graphicspath{ {images/} }

\section{Conclusion}
In conclusion, we have presented the theory and simulation of MM-HOMI and  compared them with MSI.
We confirm  that both interferences  are determined by two factors: the envelope factor and the details factor.
For MM-HOMI (MSI), the envelope factor is determined by each mode  (slit), while the details factor is determined by the N modes (slits).
The mapping relationship between MM-HOMI and MSI may provide an intuitive explanation of  MM-HOMI.
As an example of its application, we demonstrate that the square root of maximal Fisher information in a MM-HOMI increases linearly with the increase of mode numbers, indicating that increasing the mode number is a potent method for enhancing precision in quantum metrology.
%
%

\clearpage

\section*{Appendix}

\section*{A1:The calculation of multi-mode HOM interference}

In this section, we deduce the equation of HOMI using a biphoton state with a multi-mode distribution.
The joint spectral amplitude (JSA) of the biphoton state from an SPDC source can be expressed as$ f\left(\omega_s,\omega_i\right)$, and the input two-photon state is 
\begin{equation}\label{eq15}
	\begin{array}{lll}
\left|\psi\right\rangle=\iint_{-\infty}^{+\infty}{d\omega_s}d\omega_if\left(\omega_s,\omega_i\right){\hat{a}}_s^\dagger\left(\omega_s\right){\hat{a}}_i^\dagger\left(\omega_i\right)\left|00\right\rangle  ,
\end{array}
\end{equation}
where the subscripts $s$ and $i$ denote the signal and idler photons, respectively, and$\ {\hat{a}}^\dagger\left(\omega\right) $is the creation operator of the signal and idler photons at angular frequency $ \omega$.
The detection field operators of detector 1 (D1) and detector 2 (D2) are
 ${\hat{E}}_1^{\left(+\right)}\left(t_1\right)=\frac{1}{\sqrt{2\pi}}\int_{0}^{\infty}{d\omega_1{\hat{a}}_1\left(\omega_1\right)e^{-i{\omega_1}{} t_1}} $ and $ \hat{E}_2^{\left(+\right)}\left(t_2\right)=\frac{1}{\sqrt{2\pi}}\int_{0}^{\infty}{d\omega_2{\hat{a}}_2\left(\omega_2\right)e^{-i\omega_2 t_2}}
$,
where subscripts 1 and 2 denote the photons detected by $ D1$ and $D2$, respectively. The transformation rule of a 50/50 beam splitter (BS) after a delay time $ \tau $ 
is $ {\hat{a}}_1\left(\omega_1\right)=\frac{1}{\sqrt2}\left[{\hat{a}}_s\left(\omega_1\right)+{\hat{a}}_i\left(\omega_1\right)e^{-i\omega_1\tau}\right]$   and ${\hat{a}}_2\left(\omega_2\right)=\frac{1}{\sqrt2}\left[{\hat{a}}_s\left(\omega_2\right)+{\hat{a}}_i\left(\omega_2\right)e^{-i\omega_2\tau}\right]$. So, we can rewrite the field operators as
\begin{equation}\label{eq16}
	\begin{array}{lll}
{\hat{E}}_1^{\left(+\right)}\left(t_1\right)=\frac{1}{\sqrt{4\pi}}\int_{-\infty}^{+\infty}{d\omega_1}\left[{\hat{a}}_s\left(\omega_1\right)e^{-i\omega_1 t_1}+{\hat{a}}_i\left(\omega_1\right)e^{-i\omega_1(t_1+\tau)}\right] ,       
\end{array}
\end{equation}
\begin{equation}\label{eq17}
	\begin{array}{lll}{\hat{E}}_2^{\left(+\right)}\left(t_2\right)=\frac{1}{\sqrt{4\pi}}\int_{-\infty}^{+\infty}{d\omega_2}\left[{\hat{a}}_s\left(\omega_2\right)e^{-i\omega_2 t_2}+{\hat{a}}_i\left(\omega_2\right)e^{-i\omega_2(t_2+\tau)}\right] .    
\end{array}
\end{equation}
As calculated in the supplementary materials of Ref. [Optica 5, 93-98 (2018)], the two-photon coincidence probability $P\left(\tau\right)$ is
\begin{equation}\label{eq18}
	\begin{array}{lll}
P\left(\tau\right)&=\iint \nolimits_{-\infty}^{+\infty} d t_1d t_2\langle\psi \vert {\hat{E}}_1^{(-)}{\hat{E}}_2^{(-)}{\hat{E}}_2^{(+)}{\hat{E}}_1^{\left(+\right)}\vert\psi\rangle\\
          &=\frac{1}{4}\iint\nolimits_{-\infty}^{+\infty}{d\omega_1d\omega_2\left|f\left(\omega_1,\omega_2\right)-f\left(\omega_2,\omega_1\right)e^{-i\left(\omega_1-\omega_2\right)\tau}\right|^2} . 
\end{array}
\end{equation}

For simplicity, we can consider  $f\left(\omega_1,\omega_2\right)$  to be real and normalized, i.e., $f\left(\omega_1,\omega_2\right)=f^*\left(\omega_1,\omega_2\right)$ and $\iint_{-\infty}^{\infty}{d\omega_1 d\omega_2 \left|f\left(\omega_1,\omega_2\right)\right|^2 }=1$, then
\begin{equation}\label{eq19}
	\begin{array}{lll}  
	
P\left(\tau \right) = \frac{1}{2}-\frac{1}{2}\iint\nolimits_{-\infty}^{+\infty}{d\omega_1 d\omega_2}\left[f\left(\omega_2,\omega_1\right) f\left(\omega_1,\omega_2\right) \cos((\omega_1-\omega_2)\tau)\right]. 

\end{array}
\end{equation}
In the MM-HOMI,  a multi-mode JSA can be written as: 
\begin{equation}\label{eq20}
\begin{array}{l}
f\left( {{\omega _1},{\omega _2}} \right) = \sum\limits_{k = 1}^N {{f_0}\left( {{\omega _1} - {\omega _0} - \left( {2k - N - 1} \right)\alpha ,{\omega _2} - {\omega _0} + \left( {2k - N - 1} \right)\alpha } \right)} ,\\ 
\end{array}
 \end{equation} 
where $f_0\left({\omega _1},{\omega _2}\right)$ is an arbitrary distribution function of the single mode, $N$ is the mode number, 
$ \alpha $ represents the mode spacing, and $\omega _0$ is the mode's central frequency. Then 
\begin{equation}\label{eq21}
\begin{array}{lll}
 P\left( \tau  \right)& = \frac{1}{2} - \frac{1}{2}\int {\int_0^\infty  {d{\omega _1}} } d{\omega _2}\\
 &\left[{( {\sum\limits_{k = 1}^N {{f_0}\left( {{\omega _1} - {\omega _0} - \left( {2k - N - 1} \right)\alpha ,{\omega _2} - {\omega _0} + \left( {2k - N - 1} \right)\alpha } \right)} } )^2}\cos({\omega _1} - {\omega _2})\tau \right].

\end{array}
 \end{equation}
If the mode width is much smaller than the mode spacing, the cross terms can be ignored, then
 \begin{equation}\label{eq22}
\begin{array}{lll}
 &P\left( \tau  \right) \\
 &= \frac{1}{2}- \frac{1}{{2N}} \sum\limits_{k = 1}^N {\int {\int_{ - \infty }^\infty  {d{\omega _1}} } d{\omega _2}{f_0}^2\left( {{\omega _1} - {\omega _0} - \left( {2k - N - 1} \right)\alpha ,{\omega _2} - {\omega _0} + \left( {2k - N - 1} \right)\alpha } \right)} \\
  & \quad \times \cos \left( {({\omega _1} - {\omega _2})\tau } \right) \\
 &=\frac{1}{2}- \frac{1}{{2N}} \sum\limits_{k = 1}^N {\int {\int_{ - \infty }^\infty  {d{\omega _1}} } d{\omega _2}{f_0}^2\left( {{\omega _1},{\omega _2}} \right)\cos \left( {\left( {{\omega _1} - {\omega _2}} \right)\tau  + 2\left( {2k - N - 1} \right)\alpha \tau } \right)} \\
&=\frac{1}{2}- \frac{1}{2N}\sum\limits_{k = 1}^N \int\int_{- \infty}  ^\infty  {d\omega _1 } d{\omega _2}{f_0}^2\left(\omega _1,\omega _2 \right) ( \cos \left( \left( \omega _1 - \omega _2 \right)\tau  \right)\cos \left( 2\left( 2k - N - 1 \right)\alpha \tau  \right) \\
& \quad- \sin \left( \left( \omega _1 - \omega _2 \right)\tau \right )\sin \left( 2\left( 2k - N - 1 \right)\alpha \tau \right) )\\ 
 &= \frac{1}{2}\left[ 1- \frac{1}{{N}} \sum\limits_{k = 1}^N {\cos \left( {2\left( {2k - N - 1} \right)\alpha \tau } \right)\int {\int_{ - \infty }^\infty  {d{\omega _1}} } d{\omega _2}{f_0}^2\left( {{\omega _1},{\omega _2}} \right)\cos \left( {\left( {{\omega _1} - {\omega _2}} \right)\tau } \right)}\right] \\
 &=\frac{1}{2}\left[ 1- \frac{1}{{N}} \frac{{\sin (2N\alpha \tau )}}{{\sin (2\alpha \tau )}}\int {\int_{ - \infty }^\infty  {d{\omega _1}} } d{\omega _2}{f_0}^2\left( {{\omega _1},{\omega _2}} \right)\cos \left( {\left( {{\omega _1} - {\omega _2}} \right)\tau } \right)\right]\\

 &=\frac{1}{2}\left[1 - P_0 \frac{{\sin\left(  Nx  \right)}}{{\sin\left( x \right)}} \right],

\end{array}
 \end{equation}
where $x=2 \alpha\tau$, which  corresponds to the phase spacing caused by two adjacent spectral modes. 
$P_0=\frac{1}{{N}}\int {\int_{ - \infty }^\infty  {d{\omega _1}} } d{\omega _2}{f_0}^2\left( {{\omega _1},{\omega _2}} \right)\cos \left( {\left( {{\omega _1} - {\omega _2}} \right)\tau } \right)$ 
corresponds to the envelope of the interference patterns.
$P_0$ can also be written in the form of a Fourier transform:
\begin{equation}\label{eq23}
\begin{array}{lll}
P_0 &=\frac{1}{N} \int {\int_{ - \infty }^\infty  {d{\omega _1}} } d{\omega _2}{f_0}^2\left( {{\omega _1},{\omega _2}} \right)\cos \left( {\left( {{\omega _1} - {\omega _2}} \right)\tau } \right)\\
 &=  - \frac{1}{2N}\int {\int_{ - \infty }^\infty  {d{\omega _ + }} } d{\omega _ - }{f_0}^2(\frac{1}{2}\left( {{\omega _ + } + {\omega _ - }} \right),\frac{1}{2}\left( {{\omega _ + } - {\omega _ - }} \right))\cos \left( {{\omega _ - }\tau } \right)\\
 &= \int_{ - \infty }^\infty  {d{\omega _ - }} {F_0}\left( {{\omega _ - }} \right)\cos \left( {{\omega _ - }\tau } \right)\\
 &= {\mathop{\rm Re}\nolimits} \left( {\int_{ - \infty }^\infty  {d{\omega _ - }} {F_0}\left( {{\omega _ - }} \right){e^{-i{\omega _ - }\tau }}} \right)\\
 &={\mathop{\rm Re}\nolimits} (\mathcal{F} [{F_0}({\omega _ - })]),
\end{array}
 \end{equation} 
where ${\omega _ + } = {\omega _1} + {\omega _2}$, ${\omega _ - } = {\omega _1} - {\omega _2}$,  $F_0(\omega _-) =- \frac{1}{2N} \int_{ - \infty }^\infty  {d{\omega _ + }} f_0^2\left({\omega _ + },{\omega _ - }\right)$, and Re denotes the real part.

In this study, we only consider the simplest case; therefore, we assume $f\left(\omega_1,\omega_2\right) $ as a multi-mode Gaussian function \cite{Morrison_2022, Zhu2023}:
\begin{equation}\label{eq24}
	\begin{array}{lll}
f\left(\omega_1,\omega_2\right)=\sum\limits_{k=1}^{N }\exp \left[-\frac{\left(\omega_1-\omega_0-\left(2k-N-1\right)\alpha\right)^2}{\gamma^2}-\frac{\left(\omega_2-\omega_0+\left(2k-N-1\right)\alpha\right)^2}{\gamma^2}\right],         
\end{array}
\end{equation}
where $\omega_0$ is the center frequency and $\gamma$ is the mode width. Then
\begin{equation}\label{eq25}
	\begin{array}{lll}
P\left(\tau\right)&=\frac{1}{2}-\frac{1}{2}\iint \nolimits_{-\infty}^{+\infty}{d\omega_1 d\omega_2 } 
\left\{\sum\limits_{k=1}^{N }  \exp{\left[-(\frac{(\omega_1-\omega_0-(2k-N-1)\alpha)^2}{\gamma^2}-\frac{(\omega_2-\omega_0+(2k-N-1)\alpha)^2}{\gamma^2}\right]}\right\}^2
\\
&\quad\times\cos \left(\left(\omega_1-\omega_2\right)\tau\right)  .     
\end{array}
\end{equation}
If $\gamma\ll \alpha$, the cross terms can be ignored, then
\begin{equation}\label{eq26}
	\begin{array}{lll}
P\left(\tau\right)&=\frac{1}{2}-\frac{1}{2}\iint \nolimits_{-\infty}^{+\infty} {d\omega_1d\omega_2} \sum\limits_{k=1}^{N}\exp\left[-2\frac{(\omega_1-\omega_0-(2k-N-1)\alpha)^2}{\gamma^2}-2\frac{(\omega_2-\omega_0+(2k-N-1)\alpha)^2}{\gamma^2}\right]\\ 

&\quad \times\cos\left( \left(\omega_1-\omega_2\right)\tau\right) \\ 

& =\frac{1}{2}-\frac{1}{2N} \exp \left[-\frac{\gamma^2\tau^2}{4}\right] \sum\limits_{k=1}^{N}\cos \left(2\left(N+1-2k\right)\alpha\tau\right)\\ 

&=\frac{1}{2}-\frac{1}{2N} 
\exp \left[-\frac{\gamma^2\tau^2}{4}\right]\frac{\sin\left(2N\alpha\tau\right) }{\sin\left(2\alpha\tau\right)}.                    
\end{array}
\end{equation}

To understand the omission of the cross terms, we can consider the case of  N=2:
\begin{small}
\begin{equation}\label{eq27}
\begin{array}{l}
\left[{\exp{\left(- \frac{\left(\omega_1-\omega_0+\alpha\right)^2}{\gamma^2}-\frac{\left(\omega_2-\omega_0+\alpha\right)^2}{\gamma^2}\right)} }  { +\exp{\left(- \frac{\left(\omega_1-\omega_0-\alpha\right)^2}{\gamma^2}-\frac{\left(\omega_2-\omega_0-\alpha\right)^2}{\gamma^2}\right)}}\right]^2 \\  
= {\exp{\left(-2\frac{\left(\omega_1-\omega_0+\alpha\right)^2}{\gamma^2}-2\frac{ \left(\omega_2-\omega_0+\alpha\right)^2}{\gamma^2}\right)} }  { +\exp{\left(-2\frac{\left(\omega_1-\omega_0-\alpha\right)^2}{\gamma^2}-2\frac{\left(\omega_2-\omega_0-\alpha\right)^2}{\gamma^2}\right)}}\\
\quad+ 2\exp{\left(-\frac{\left(\omega_1-\omega_0+\alpha\right)^2}{\gamma^2}-\frac{\left(\omega_2-\omega_0+\alpha\right)^2}{\gamma^2}- \frac{\left(\omega_1-\omega_0-\alpha\right)^2}{\gamma^2}-\frac{\left(\omega_2-\omega_0-\alpha\right)^2}{\gamma^2} \right)} .
\end{array}
\end{equation}
\end{small}
The last term is the cross-term. 
When $\gamma\ll \alpha$, the cross-term is much smaller than the sum of the first and second terms, so it can be ignored.

\section*{A2: The calculation of multi-slit interference} 

As shown in Fig.\,\ref{fig:1}(b), the number of slits is N, the separation between adjacent slits is $d$, and the width of each slit is $a$. The diffraction angle associated with  point $p$ is${\rm{\;}}\theta $. The optical path difference between two adjacent slits is $ d\sin\theta $ and the corresponding phase difference is  $ 2\pi d\sin\theta/ \lambda $. $\lambda$ is the wavelength of the incident monochromatic light. 

Firstly, the light intensity distribution of single-slit diffraction is calculated according to the Fresnel-Kirchhoff diffraction integral formula.
The light diffracted from one direction is focused on the focal plane of the lens L at  point $p$.
 The amplitude of single-slit diffraction is 
\begin{equation}\label{eq28}
\begin{array}{lll}
A_1 &= C\int_{ - \infty}^{\infty} dx{{f_0\left(x\right)e^{ - ikx\sin \theta }}} 
=C\int_{ - \infty}^{\infty} dx{{f_0\left(x\right)e^{ - i2\pi x\frac{{\sin \theta }}{\lambda } }}} 
=C\mathcal{F} \left[{f_0}\left(x \right)\right],
\end{array}
\end{equation}
where $k = \frac{{2\pi }}{\lambda }$, $f_0\left(x\right)$ is aperture function, and $C$ is a constant, which is determined by the power of the light source, the distance between the slit and the screen, and the size of the slit \cite{born1980}. Intensity of the light is the magnitude of the Fourier Transform of aperture function. In this study, we only consider the simplest case. We set aperture function$f_0(x)$ as rectangle function:
 
\begin{equation}\label{eq29}
\begin{array}{lll}
A_1 &= C\int_{ - \frac{a}{2}}^{\frac{a}{2}} {{e^{ - ikx\sin \theta }}dx} 
= \frac{{2C}}{{k\sin \theta }}\sin \left( {\frac{{ka\sin \theta }}{2}} \right).
\end{array}
\end{equation}
 Let $u = \frac{{\pi a\sin\theta }}{\lambda }$, then
\begin{equation}\label{eq30}
\begin{array}{lll}
A_1 = aC\frac{{\sin u}}{u},
\end{array}
\end{equation}
when $\theta=0$, $u=0, \frac{{\sin u}}{u}=1$,  the amplitude due to one slit is $A_0 = aC$. So the amplitude of single-slit diffraction is
\begin{equation}\label{eq31}
\begin{array}{lll}
A_1 = A_0\frac{{\sin u}}{u}.
\end{array}
\end{equation}
After passing through the first slit, the light field  at point $p$ is 
\begin{equation}\label{eq32}
	\begin{array}{lll}
	E_1=A_1 \cos\left(\omega t + \varphi_0\right).
\end{array}
\end{equation}

The optical path difference between the two adjacent slits is equal, so the phase difference is equal. Therefore, the general equation of the light field at point $p$ is
\begin{equation}\label{eq33}
\begin{array}{lll}
    E_i=A_1 \cos\left(\omega t + \varphi_0+\left(i-1\right)\delta \right)
	      \quad\left(i=1,2,\cdots N\right),

\end{array}
\end{equation}
where $\delta=\frac{2\pi d \sin\theta}{\lambda} $ is the phase difference between the two adjacent slits. $i$ is the number of the slit.
Therefore, the combined field at point $p$ is
\begin{equation}\label{eq34}
	\begin{array}{lll}

    E\left(p,t\right)&=\sum\limits_{i=1}^{N}{E_i=}A_1\sum\limits_{i=1}^{N}{\cos\left[\omega t+\varphi_0+\left(i-1\right)\delta\right]} \\  
    
 &= \frac{{{A_1}}}{{2\sin (\frac{\delta }{2})}}\sum\limits_{i = 1}^N {\cos \left[\omega t + {\varphi _0} + \left(i - 1\right)\delta \right]\sin\left (\frac{\delta }{2}\right)} \\  

   & =\frac{A_1}{2\sin(\frac{\delta}{2})}\sum\limits_{i=1}^{N} \sin\left[\omega t+\varphi_0+\left(i-\frac{1}{2}\right)\delta\right]-\sin\left[\omega t+\varphi_0+\left(i-\frac{3}{2}\right)\delta\right]\\  
   &=\frac{A_1 \sin\left(\frac{N\delta}{2}\right)}{\sin\left(\frac{\delta}{2}\right)}\cos\left[\omega t+\varphi_0+\frac{N-1}{2}\delta\right].
\end{array}
\end{equation}
Finally,  the amplitude is
\begin{equation}\label{eq35}
	\begin{array}{lll}
A_p=\frac{A_1 \sin(\frac{N\delta}{2})}{\sin\left(\frac{\delta}{2}\right)}={A_0}\frac{{\sin u}}{u}\frac{{\sin\left( {Nv} \right)}}{{\sin\left( v \right)}},
\end{array}
\end{equation}
 where  $u = \frac{{\pi a\sin\theta }}{\lambda }$, $v = \frac{{\pi d\sin\theta }}{\lambda }=\frac{\delta}{2}$.
The intensity of the diffraction pattern is 
\begin{equation}
\label{eq36}
\begin{array}{lll}
I=A_p A_p^\ast
=\left(A_0\right)^2\left(\frac{{\sin}{\left(u\right)}}{u}\right)^2 \left(\frac{\sin{\left(N v\right)}}{\sin\left(v\right)}\right)^2.\ \ \ 
\end{array}
\end{equation}

\section*{A3:The calculation of multi-mode Mach–Zehnder interference}

In this section, we consider the interference pattern in a multi-mode Mach–Zehnder interference, with a setup shown in Fig.\,\ref{fig:7}(a). As calculated in detail in the supplementary materials of Ref. \cite{Jin2018Optica}, the coincidence probability $P_{MZI}\left(\tau\right)$ as a function of optical path delay $\tau$ can be expressed as
\begin{equation}\label{eq37}
\begin{array}{lll}
P_{MZI}\left( \tau  \right) = \frac{1}{2}\left[ 1  + \int\nolimits_{ - \infty }^{ + \infty } {d\omega {{\left| {f\left( \omega  \right)} \right|}^2}\cos \left( {\omega \tau } \right)}  \right],
\end{array}
\end{equation}
where $f\left(\omega\right)$ is the spectral amplitude of the laser source. For simplicity,   we have assumed  that $f\left(\omega\right)$  is normalized in the above equation, i.e.,  $\int_{-\infty}^{\infty}{d\omega} |f\left(\omega\right)|^2 =1$.
In multi-mode Mach-Zehnder interference, $f(\omega)$ can be written as: 
\begin{equation}\label{eq38}
\begin{array}{l}
f\left( {\omega } \right) = \sum\limits_{k = 1}^N {{f_0}\left( {{\omega} - {\omega _0} - \left( {2k - N - 1} \right)\alpha  } \right)} ,\\ 
\end{array}
 \end{equation} 
where $f_0\left(\omega\right)$ is an arbitrary distribution function of the single mode, $N$ is the mode number, 
$ \alpha $ represents the mode spacing, and $\omega _0$ is the mode's central frequency. If the mode width is much smaller than the mode spacing, the cross terms can be ignored, 
then 
\begin{equation}\label{eq39}
\begin{array}{lll}
P_{MZI}\left( \tau  \right)&=\frac{1}{2}  +\frac{1}{2} \int\nolimits_{ - \infty }^{ + \infty } d\omega {\left| { \sum\limits_{k = 1}^N {{f_0}\left( {{\omega} - {\omega _0} - \left( {2k - N - 1} \right)\alpha  } \right)}} \right|^2}\cos \left( {\omega \tau } \right) \\  
&=\frac{1}{2}  +\frac{1}{2} \int\nolimits_{ - \infty }^{ + \infty } d\omega {\left|  \sum\limits_{k = 1}^N {{f_0}\left( \omega\right)} \right|^2}\cos\left ( {\left(\omega +{\omega _0} + \left( {2k - N - 1} \right)\alpha  \right)}\tau  \right) \\  
 &= \frac{1}{2}\left[ 1-  \sum\limits_{k = 1}^N {\cos\left ( \omega_0\tau+{\left( {2k - N - 1} \right)\alpha \tau } \right)} {\int_{ - \infty }^\infty  {d\omega} } \left|{f_0}\left( \omega  \right)\right|^2\cos \left( \left( \omega \right)\tau  \right)\right] \\
 &=\frac{1}{2}\left[ 1- \frac{1}{{N}} \frac{{\sin \left(N\alpha \tau\right )}}{{\sin\left (\alpha \tau \right)}}\cos{\left(\omega_0\tau\right)} {\int_{ - \infty }^\infty  {d{\omega}} } \left|{f_0}\left( {\omega } \right)\right|^2\cos ( \omega \tau )\right]\\

 &=\frac{1}{2}\left[1 - P_0 \frac{{\sin\left(  Nx  \right)}}{{\sin\left( x \right)}} \cos{\left(\omega_0\tau\right)}\right],
 
\end{array}
\end{equation}
where $x= \alpha\tau$ and $P_0=\frac{1}{{N}} {\int_{ - \infty }^\infty  {d{\omega}} } \left|{f_0}\left( {{\omega }} \right)\right|^2\cos \left( {{\omega  } \tau } \right)$ 
corresponds to the envelope of the interference patterns.
 
To simplify the analysis,  we assume that $f\left(\omega\right) $ is a multi-mode Gaussian function:
\begin{equation}\label{eq40}
\begin{array}{lll}
f\left( \omega  \right) = \sum\limits_{k = 1}^N {\exp \left[ { - \frac{{{{\left( {\omega  - {\omega _0} - \left( {2k - N - 1} \right)\alpha } \right)}^2}}}{{{\gamma ^2}}}} \right]} ,
\end{array}
\end{equation}
where $\omega_0$ is the center frequency and $\alpha$ is the mode spacing. Then
\begin{equation}\label{eq41}
\begin{array}{lll}
P_{MZI}\left( \tau  \right)&=\frac{1}{2}  +\frac{1}{2} \int\nolimits_{ - \infty }^{ + \infty } {d\omega {{\left| {f\left( \omega  \right)} \right|}^2}\cos \left( {\omega \tau } \right)} \\  
& =  \frac{1}{2} + \frac{1}{2} {\exp \left[ { - \frac{{{\gamma ^2}{\tau ^2}}}{8}} \right]\sum\limits_{k = 1}^N {\cos \left( {{\omega _0}\tau  + \left( {2k - N - 1} \right)\alpha \tau } \right)} } \\ 
 &= \frac{1}{2} + \frac{1}{{2N}}\exp \left[ { - \frac{{{\gamma ^2}{\tau ^2}}}{8}} \right]\frac{{\sin \left( {N\alpha \tau } \right)}}{{\sin \left( {\alpha \tau } \right)}}\cos \left( {{\omega _0}\tau } \right)\\  
&= \frac{1}{2} + \frac{1}{{2N}}\exp \left[ { - \frac{{{\gamma ^2}{\tau ^2}}}{8}} \right]\frac{{\sin \left( {Nx} \right)}}{{\sin \left( {x} \right)}}\cos \left( {{\omega _0}\tau } \right),
\end{array}
\end{equation}
where $x= \alpha\tau$.

\section*{A4: The calculation of multi-mode NOON-state interference}

In this section, we consider the interference pattern in a multi-mode NOON-state interference, with a setup shown in Fig.\,\ref{fig:7}(b).  Here, the NOON-state is a 
$(\left| {20} \right\rangle  + \left| {02} \right\rangle )/\sqrt 2$ 
state, which has a photon number of 2, but with a spectral-mode number of N, as shown in  Fig.\,\ref{fig:7}(e1-e8).
As calculated in detail in the supplementary materials of Ref.\cite{Jin2018Optica} , the coincidence probability $P_{NOON}\left(\tau\right)$ as a function of optical path delay $\tau$ can be expressed as
\begin{equation}\label{eq42}
\begin{array}{lll}
P_{NOON}\left(\tau \right) = \frac{1}{2} + \frac{1}{2} \mathop \int\!\!\!\int \nolimits_{-\infty}^{\infty}  d{\omega _1}d{\omega _2} {\left|f\left( {{\omega _1},{\omega _2}} \right)\right|^2\cos\left ({\omega _1} + {\omega _2}\right)\tau } ,
\end{array}
\end{equation}
where $f\left(\omega_1,\omega_2\right)$ is the joint spectral amplitude of the biphoton.  For simplicity,  in the preceding equation we has assumed  $f\left(\omega_1,\omega_2\right)$  normalized, i.e., $\iint_{-\infty}^{\infty}{d\omega_1 d\omega_2 |f\left(\omega_1,\omega_2\right)|^2 }=1$, and  $f\left(\omega_1,\omega_2\right)$  satisfies the exchanging symmetry of   $f\left(\omega_1,\omega_2\right)= f\left(\omega_2,\omega_1\right)$.

For simplicity, we can further set  $f\left(\omega_1,\omega_2\right)$  as real and normalized, i.e., $f\left(\omega_1,\omega_2\right)=f^*\left(\omega_1,\omega_2\right)$ and $\iint_{-\infty}^{\infty}{d\omega_1 d\omega_2 \left|f\left(\omega_2,\omega_1\right)\right|^2 }=1$. 
In the multi-mode NOON-state interference , $f\left(\omega_1,\omega_2\right)$ can also be written as: 
\begin{equation}\label{eq43}
\begin{array}{l}
f\left( {{\omega _1},{\omega _2}} \right) = \sum\limits_{k = 1}^N {{f_0}\left( {{\omega _1} - {\omega _0} - \left( {2k - N - 1} \right)\alpha ,{\omega _2} - {\omega _0} - \left( {2k - N - 1} \right)\alpha } \right)} ,\\ 
\end{array}
 \end{equation} 
where $f_0\left({\omega _1},{\omega _2}\right)$ is an arbitrary distribution function of the single mode, $N$ is the mode number, 
$ \alpha $ represents the mode spacing, and $\omega _0$ is the mode's central frequency. If the mode width is much smaller than the mode spacing, the cross terms can be ignored,
then
\begin{equation}\label{eq44}
	\begin{array}{lll}
P_{NOON}\left( \tau  \right)& = \frac{1}{2} + \frac{1}{2}\mathop \int\!\!\!\int \nolimits_{ - \infty }^\infty d{\omega _1}d{\omega _2}\cos \left(\left({\omega _1} + {\omega _2}\right)\tau\right) \\
&\quad\times \left| \sum\limits_{k = 1}^N f_0\left( {\omega _1} - {\omega _0} - \left( {2k - N - 1} \right)\alpha ,{\omega _2} - {\omega _0} - \left( {2k - N - 1} \right)\alpha  \right) \right|^2\\ 
& = \frac{1}{2} + \frac{1}{2}\sum\limits_{k = 1}^N\mathop \int\!\!\!\int \nolimits_{ - \infty }^\infty d{\omega _1}d{\omega _2}  \left| {f_0}\left( \omega _1  ,\omega _2  \right)\right|^2\\
&\quad\times\cos\left(\left({\omega _1} + {\omega _2}+2\omega _0+2\left( {2k - N - 1} \right)\alpha\right)\tau\right) \\ 
 &= \frac{1}{2} + \frac{1}{{2N}}\sum\limits_{k = 1}^N {\cos \left( {2\tau \left( {{\omega _0} + (2k - N - 1)\alpha } \right)} \right)}\\
 &\quad\times\int\!\!\!\int \nolimits_{ - \infty }^\infty d{\omega _1}d{\omega _2}  \left| {f_0}\left( \omega _1  ,\omega _2  \right)\right|^2\cos\left(\omega _1+\omega _2\right)\\   
 
 & =\frac{1}{2} + \frac{1}{{2N}}\frac{{\sin \left( {Nx } \right)}}{{\sin \left( {x } \right)}}\cos \left( {2{\omega _0}\tau } \right)\int\!\!\!\int \nolimits_{ - \infty }^\infty d{\omega _1}d{\omega _2} \left | {f_0}\left( \omega _1  ,\omega _2  \right)\right|^2\cos\left(\omega _1+\omega _2\right)\\
 & =\frac{1}{2} + \frac{1}{{2}} P_0\frac{{\sin \left( {Nx } \right)}}{{\sin \left( {x } \right)}}\cos \left( {2{\omega _0}\tau } \right),
\end{array}
\end{equation}
where $x=2 \alpha\tau$ and $P_0=\frac{1}{{N}}\int {\int_{ - \infty }^\infty  {d{\omega _1}} } d{\omega _2}\left|{f_0}\left( {{\omega _1},{\omega _2}} \right)\right|^2\cos \left( {\left( {{\omega _1} + {\omega _2}} \right)\tau } \right)$ 
corresponds to the envelope of the interference patterns.
For simplicity, we also assume $f\left(\omega_1,\omega_2\right) $ as a multi-mode Gaussian function:
\begin{equation}\label{eq45}
	\begin{array}{lll}
f\left(\omega_1,\omega_2\right)=\sum\limits_{k=1}^{N }\exp\left[-\frac{\left(\omega_1-\omega_0-\left(2k-N-1\right)\alpha\right)^2}{\gamma^2}-\frac{\left(\omega_2-\omega_0+\left(2k-N-1\right)\alpha\right)^2}{\gamma^2}\right],         
\end{array}
\end{equation}
where $\omega_0$ is the center frequency and $\alpha$ is the mode separation.
Then,
\begin{equation}\label{eq46}
	\begin{array}{lll}
P_{NOON}\left( \tau  \right)& = \frac{1}{2} + \frac{1}{2}\mathop \int\!\!\!\int \nolimits_{ - \infty }^\infty d{\omega _1}d{\omega _2}\cos (({\omega _1} + {\omega _2})\tau)  \\
&\quad\times\sum\limits_{k = 1}^N {\exp } \left[ - 2\frac{{{{\left({\omega _1} - {\omega _0} -\left (2k - N - 1\right)\alpha \right)}^2}}}{{{\gamma ^2}}} - 2\frac{{{{\left({\omega _2} - {\omega _0} - \left(2k - N - 1\right)\alpha \right)}^2}}}{{{\gamma ^2}}}\right]\\    
 &= \frac{1}{2} + \frac{1}{{2N}}\exp \left[ - \frac{{{\gamma ^2}{\tau ^2}}}{4}\right]\sum\limits_{k = 1}^N {\cos \left( {2\tau \left( {{\omega _0} + (2k - N - 1)\alpha } \right)} \right)} \\   
 & = \frac{1}{2} + \frac{1}{{2N}}\exp \left[ - \frac{{{\gamma ^2}{\tau ^2}}}{4}\right]\frac{{\sin \left( {2N\alpha \tau } \right)}}{{\sin \left( {2\alpha \tau } \right)}}\cos \left( {2{\omega _0}\tau } \right)\\  
 & =\frac{1}{2} + \frac{1}{{2N}}\exp \left[ - \frac{{{\gamma ^2}{\tau ^2}}}{4}\right]\frac{{\sin \left( {Nx } \right)}}{{\sin \left( {x } \right)}}\cos \left( {2{\omega _0}\tau } \right),
\end{array}
\end{equation}
where $x=2 \alpha\tau$,  which  determines the phase spacing caused by two adjacent spectral modes.

\section*{Funding}
This work was supported by the National Natural Science Foundation of China (Grant Numbers 92365106, 12074299, and 11704290)  and the Natural Science Foundation of Hubei Province (2022CFA039).
\section*{Disclosures}
The authors declare no conflicts of interest.
\section*{Data Availability}
Data underlying the results presented in this paper are not publicly available at this time but may be obtained from the authors upon reasonable request.

\bibliography{multi1}

\begin{thebibliography}{10}
\newcommand{\enquote}[1]{``#1''}

\bibitem{Hong1987}
C.~K. Hong, Z.~Y. Ou, and L.~Mandel, \enquote{Measurement of subpicosecond time
  intervals between two photons by interference,} {\protect\JournalTitle{Phys.
  Rev. Lett.}} \textbf{59}, 2044--2046 (1987).

\bibitem{AgataM2017}
A.~M. Bra\'{n}czyk, \enquote{{Hong-Ou-Mandel } interference,}
  {\protect\JournalTitle{arXiv:1711.00080}}  (2017).

\bibitem{Bouchard2020}
F.~Bouchard, A.~Sit, Y.~Zhang, R.~Fickler, F.~M. Miatto, Y.~Yao, F.~Sciarrino,
  and E.~Karimi, \enquote{Two-photon interference: the {Hong-Ou-Mandel}
  effect,} {\protect\JournalTitle{Rep. Prog. Phys.}} \textbf{84}, 012402
  (2020).

\bibitem{Liu_2021}
Y.~Liu, R.~Quan, X.~Xiang, H.~Hong, M.~Cao, T.~Liu, R.~Dong, and S.~Zhang,
  \enquote{Quantum clock synchronization over 20-km multiple segmented fibers
  with frequency-correlated photon pairs and {HOM} interference,}
  {\protect\JournalTitle{Appl. Phys. Lett.}} \textbf{119}, 144003 (2021).

\bibitem{Yang_2022}
C.~Yang, S.-J. Niu, Z.-Y. Zhou, Y.~Li, Y.-H. Li, Z.~Ge, M.-Y. Gao, Z.-Q.-Z.
  Han, R.-H. Chen, G.-C. Guo, and B.-S. Shi, \enquote{Advantages of the
  frequency-conversion technique in quantum interference,}
  {\protect\JournalTitle{Phys. Rev. Appl.}} \textbf{105}, 063715 (2022).

\bibitem{Jin2016QST}
R.-B. Jin, R.~Shimizu, M.~Fujiwara, M.~Takeoka, R.~Wakabayashi, T.~Yamashita,
  S.~Miki, H.~Terai, T.~Gerrits, and M.~Sasaki, \enquote{Simple method of
  generating and distributing frequency-entangled qudits,}
  {\protect\JournalTitle{Quantum Sci. Technol.}} \textbf{1}, 015004 (2016).

\bibitem{Useche2021}
D.~H. Useche, A.~Giraldo-Carvajal, H.~M. Zuluaga-Bucheli, J.~A.
  Jaramillo-Villegas, and F.~A. Gonz{\'{a}}lez, \enquote{Quantum measurement
  classification with qudits,} {\protect\JournalTitle{Quantum Inform.
  Process.}} \textbf{21} (2021).

\bibitem{Castro2022}
A.~Castro, A.~Garc\'{\i}a~Carrizo, S.~Roca, D.~Zueco, and F.~Luis,
  \enquote{Optimal control of molecular spin qudits,}
  {\protect\JournalTitle{Phys. Rev. Appl.}} \textbf{17}, 064028 (2022).

\bibitem{Yang:23}
Z.-X. Yang, Z.-Q. Zeng, Y.~Tian, S.~Wang, R.~Shimizu, H.-Y. Wu, S.~Liu, and
  R.-B. Jin, \enquote{Spatial--spectral mapping to prepare frequency entangled
  qudits,} {\protect\JournalTitle{Opt. Lett.}} \textbf{48}, 2361--2364 (2023).

\bibitem{Xiang2013}
G.~Y. Xiang, H.~F. Hofmann, and G.~J. Pryde, \enquote{Optimal multi-photon
  phase sensing with a single interference fringe,} {\protect\JournalTitle{Sci.
  Rep.}} \textbf{3}, 2684--2684 (2013).

\bibitem{Jin2016SR}
R.-B. Jin, M.~Fujiwara, R.~Shimizu, R.~J. Collins, G.~S. Buller, T.~Yamashita,
  S.~Miki, H.~Terai, M.~Takeoka, and M.~Sasaki, \enquote{{Detection-dependent
  six-photon Holland-Burnett state interference},} {\protect\JournalTitle{Sci.
  Rep.}} \textbf{6}, 36914 (2016).

\bibitem{Lyons2018SA}
A.~Lyons, C.~G. Knee, E.~Bolduc, T.~Roger, J.~Leach, M.~E. Gauger, and
  D.~Faccio, \enquote{{Attosecond-resolution Hong-Ou-Mandel interferometry},}
  {\protect\JournalTitle{Sci. Adv.}} \textbf{4}, eaap9416 (2018).

\bibitem{Lingaraju2019}
N.~B. Lingaraju, H.-H. Lu, S.~Seshadri, P.~Imany, D.~E. Leaird, J.~M. Lukens,
  and A.~M. Weiner, \enquote{Quantum frequency combs and {Hong-Ou-Mandel}
  interferometry: the role of spectral phase coherence,}
  {\protect\JournalTitle{Opt. Express}} \textbf{27}, 38683--38697 (2019).

\bibitem{Chen2019njpQIursin}
Y.~Chen, M.~Fink, F.~Steinlechner, J.~P. Torres, and R.~Ursin,
  \enquote{{Hong-Ou-Mandel interferometry on a biphoton beat note},}
  {\protect\JournalTitle{npj Quantum Inform.}} \textbf{5}, 43 (2019).

\bibitem{Chen2021}
Y.~Chen, S.~Ecker, L.~Chen, F.~Steinlechner, M.~Huber, and R.~Ursin,
  \enquote{Temporal distinguishability in {Hong-Ou-Mandel} interference for
  harnessing high-dimensional frequency entanglement,}
  {\protect\JournalTitle{npj Quantum Inform.}} \textbf{7}, 167 (2021).

\bibitem{Morrison_2022}
C.~L. Morrison, F.~Graffitti, P.~Barrow, A.~Pickston, J.~Ho, and A.~Fedrizzi,
  \enquote{Frequency-bin entanglement from domain-engineered down-conversion,}
  {\protect\JournalTitle{{APL} Photon.}} \textbf{7}, 066102 (2022).

\bibitem{Xue_2019}
R.~Xue, X.~Yao, X.~Liu, H.~Wang, H.~Li, Z.~Wang, L.~You, Y.~Huang, and
  W.~Zhang, \enquote{Spatial quantum beating of adjustable biphoton frequency
  comb with high-dimensional frequency-bin entanglement,}
  {\protect\JournalTitle{{IEEE} Photon. J.}} \textbf{11}, 1--9 (2019).

\bibitem{Fabre2020}
N.~Fabre, G.~Maltese, F.~Appas, S.~Felicetti, A.~Ketterer, A.~Keller,
  T.~Coudreau, F.~Baboux, M.~I. Amanti, S.~Ducci, and P.~Milman,
  \enquote{Generation of a time-frequency grid state with integrated biphoton
  frequency combs,} {\protect\JournalTitle{{Phys. Rev. A}}} \textbf{102},
  012607 (2020).

\bibitem{Xie_2015}
Z.~Xie, T.~Zhong, S.~Shrestha, X.~Xu, J.~Liang, Y.-X. Gong, J.~C. Bienfang,
  A.~Restelli, J.~H. Shapiro, F.~N.~C. Wong, and C.~W. Wong,
  \enquote{Harnessing high-dimensional hyperentanglement through a biphoton
  frequency comb,} {\protect\JournalTitle{Nat. Photon.}} \textbf{9}, 536--542
  (2015).

\bibitem{Chang_2021}
K.-C. Chang, X.~Cheng, M.~C. Sarihan, A.~K. Vinod, Y.~S. Lee, T.~Zhong, Y.-X.
  Gong, Z.~Xie, J.~H. Shapiro, F.~N.~C. Wong, and C.~W. Wong, \enquote{648
  hilbert-space dimensionality in a biphoton frequency comb: entanglement of
  formation and schmidt mode decomposition,} {\protect\JournalTitle{npj Quantum
  Inform.}} \textbf{7} (2021).

\bibitem{born1980}
M.~Born and E.~Wolf, \emph{Principles of Optics, 6-th ed} (Cambridge
  University, 1980).

\bibitem{Young1807}
T.~Young, \enquote{A course of lectures on natural philosophy and the
  mechanical arts,} {\protect\JournalTitle{Lecture 39}} \textbf{1}, 463--464
  (1807).

\bibitem{Hariharan2003}
P.~Hariharan, \emph{Optical interferometry} (Academic, Amsterdam Boston, 2003).

\bibitem{Jewett2008}
J.~W. Jewett and R.~Serway, \emph{Physics for Scientists and Engineers with
  Modern physics} (Pearson Education, Upper Saddle River, N.J, 2008).

\bibitem{Young2012}
H.~D. Young, R.~A. Freedman, and A.~L. Ford, \emph{Sears and Zemansky's
  University Physics with Modern Physics, 13th Edition} (Addison-Wesley,
  Boston, 2012).

\bibitem{URen2006}
A.~B. U'Ren, R.~K. Erdmann, M.~de~la Cruz-Gutierrez, and I.~A. Walmsley,
  \enquote{Generation of two-photon states with an arbitrary degree of
  entanglement via nonlinear crystal superlattices,}
  {\protect\JournalTitle{Phys. Rev. Lett.}} \textbf{97}, 223602 (2006).

\bibitem{Mosley2008}
P.~J. Mosley, J.~S. Lundeen, B.~J. Smith, P.~Wasylczyk, A.~B. U'Ren,
  C.~Silberhorn, and I.~A. Walmsley, \enquote{Heralded generation of ultrafast
  single photons in pure quantum states,} {\protect\JournalTitle{Phys. Rev.
  Lett.}} \textbf{100}, 133601 (2008).

\bibitem{LI2023OLT}
B.~Li, B.~Yuan, C.~Chen, X.~Xiang, R.~Quan, R.~Dong, S.~Zhang, and R.-B. Jin,
  \enquote{Spectrally resolved two-photon interference in a modified
  {Hong–Ou–Mandel} interferometer,} {\protect\JournalTitle{Optics \& Laser
  Technology}} \textbf{159}, 109039 (2023).

\bibitem{Grice1997}
W.~P. Grice and I.~A. Walmsley, \enquote{Spectral information and
  distinguishability in {type-II} down-conversion with a broadband pump,}
  {\protect\JournalTitle{Phys. Rev. A}} \textbf{56}, 1627--1634 (1997).

\bibitem{Jin2018Optica}
R.-B. Jin and R.~Shimizu, \enquote{Extended {Wiener-Khinchin} theorem for
  quantum spectral analysis,} {\protect\JournalTitle{Optica}} \textbf{5},
  93--98 (2018).

\bibitem{Zhu2023}
J.-L. Zhu, W.-X. Zhu, X.-T. Shi, C.-T. Zhang, X.~Hao, Z.-X. Yang, and R.-B.
  Jin, \enquote{Design of mid-infrared entangled photon sources using lithium
  niobate,} {\protect\JournalTitle{J. Opt. Soc. Am. B}} \textbf{40}, A9 (2023).

\bibitem{Cramer1999}
H.~Cram{\'e}r, \emph{Mathematical methods of statistics}, vol.~26 (Princeton
  University, 1999).

\bibitem{Jin2021arXiv}
R.-B. Jin, R.~Shimizu, T.~Ono, M.~Fujiwara, G.-W. Deng, Q.~Zhou, M.~Sasaki, and
  M.~Takeoka, \enquote{{Spectrally resolved NOON state interference},}
  {\protect\JournalTitle{arXiv: 2104.01062}}  (2021).

\end{thebibliography}

\end{document}